\documentclass[twocolumn]{aastex62}
\usepackage{graphics,epsf}
\usepackage{amsmath}                % American Mathematical Society package
\usepackage{amsfonts}               % American Mathematical Society fonts
\usepackage{amssymb}                % American Mathematical Society symbol
\usepackage{epsfig}                 % EPS figures
\usepackage{appendix}
\usepackage{graphicx}
\usepackage{float}
\usepackage{color}
\usepackage{multirow}
\usepackage{colortbl}
\usepackage[para,online,flushleft]{threeparttable}

\newcommand{\erg}{{~\rm erg}}
\newcommand{\yr}{{~\rm yr}}

\newcommand{\AU}{{~\rm AU}}

\begin{document}

\title{Parasite common envelope evolution by triple-star systems}

%\correspondingauthor{Ealeal Bear, Noam Soker}
\email{soker@physics.technion.ac.il; ealealbh@gmail.com}

\author{Noam Soker}
\affiliation{Department of Physics, Technion – Israel Institute of Technology, Haifa 3200003, Israel}
\affiliation{Guangdong Technion Israel Institute of Technology, Guangdong Province, Shantou 515069, China}

\author{Ealeal Bear}
\affiliation{Department of Physics, Technion – Israel Institute of Technology, Haifa 3200003, Israel}

\begin{abstract}
We study a scenario by which a giant wide tertiary star engulfs and forces a tight binary system of a white dwarf (WD) and a main sequence (MS) star to enter a common envelope evolution (CEE) with each other, and then unbinds the WD-MS common envelope.
The WD-MS binary system, now with the WD inside the MS envelope, does not have sufficient orbital energy to unbind their common envelope. However, as they approach the center of the giant star Roche lobe overflow to the core of the giant star and/or merger of the WD with the core remove a large fraction of the WD-MS common envelope or all of it. Namely, the energy source for unbinding the WD-MS tight common envelope is the triple-star CEE. For that we term this scenario a \textit{parasite CEE}. Overall, the destruction of the MS star absorbs energy from the triple-star system, a process that might lead to WD-core merger during the triple-star CEE. 
The parasite CEE leaves behind either one massive WD that in some cases might explode as a peculiar type Ia supernova or two close WDs that at later time might explode as a type Ia supernova. 
We very crudely estimate the rate of the parasite CEE to be a fraction of $\approx 0.001$ out of all evolved triple stars. 
\end{abstract}

\keywords{(stars:) binaries (including multiple): close; stars: AGB and post-AGB; white dwarfs;} 

% ==========================================================
\section{Introduction} 
\label{sec:intro}
% ==========================================================

Most studies of the common envelope evolution (CEE) consider the evolution of a compact object that is spiralling-in inside the extended envelope of a giant star. In that case the compact object and the core of the giant star can release a huge amount of gravitational energy as the orbit shrinks, energy that inflates the envelope and leads to the ejection of common envelope matter (CEM), processes that three-dimensional numerical simulations and other studies have demonstrated over the years (e.g., \citealt{LivioSoker1988, RasioLivio1996, Termanetal1995, SandquistTaam1998, Sandquistetal2000, Lombardi2006, RickerTaam2008, TaamRicker2010, DeMarcoetal2011, Passyetal2011, Passyetal2012, RickerTaam2012, Nandezetal2014, Ohlmannetal2016, Ohlmannetal2016b, Staffetal2016MN8, NandezIvanova2016, Kuruwitaetal2016, IvanovaNandez2016, Iaconietal2017, DeMarcoIzzard2017, Galavizetal2017, Iaconietal2018, MacLeodetal2018, Chamandyetal2020, Krameretal2020, Reichardtetal2020, Sandetal2020, GlanzPerets2021}).

The three-dimensional hydrodynamical numerical simulations also show that it is not straightforward to eject the entire CEM by the orbital energy alone. This failure to eject the entire CEM in simulations as well as other considerations, e.g., the limited amount of orbital energy that planets can deposit in CEE, brought suggestions for extra energy sources that further eject the CEM, e.g., dust formation (e.g., \citealt{Soker1992b, Soker1998AGB, GlanzPerets2018, Iaconietal2019, Iaconi2020}), recombination energy, and jets. 
The formation of large quantities of dust on the cool giant's surface can utilise the high luminosity of  red giant branch (RGB), asymptotic giant branch (AGB), and red supergiant (RSG) stars. It is still under debate (e.g., \citealt{Reichardtetal2020}) whether the recombination energy of the CEM can add much to unbind the envelope, with studies that claim for (e.g.,  \citealt{IvanovaNandez2016, Kruckowetal2016, NandezIvanova2016, Ivanova2018}) and against (e.g., \citealt{Sabachetal2017, Grichener2018, WilsonNordhaus2019}) its important role. 
A compact star that accretes mass in a CEE is likely to do so through an accretion disk that in turn is likely to launch jets. These jets inflate the envelope and deposit extra energy to facilitate envelope removal when the compact companion is a main sequence (MS) star (e.g., \citealt {ShiberSoker2018, Shiberetal2019}) or a neutron star or a black hole (e.g., \citealt{MorenoMendezetal2017, LopezCamaraetal2019, LopezCamaraetal2020MN, GrichenerSoker2021}).

The CEE of a white dwarf (WD) inside a MS star differs in several key issues from the CEE of a compact object in a giant star.   
(1) It is not so easy to set such a CEE.  Because the orbital moment of inertia of the WD-MS system is larger than the MS moment of inertia,  the WD is likely to bring the MS rotation (spin) to synchronisation (corotation) with the orbital motion, reducing to zero the tidal forces. The mass transfer process can be stable as in cataclysmic variables. (2) As we will show in section \ref{sec:BindingEnergy} a WD spiralling-in inside a MS star does not liberate sufficient orbital energy to unbind the envelope. 
(3) The  extra energy sources are negligible in this type of CEE. Recombination energy is negligible with respect to the potential well of the MS star, surface temperature is too high for dust formation, and a WD is unlikely to accrete mass at a high enough rate to launch energetic jets. The reason for the latter is that for an accretion rate of $\dot M_{\rm WD} \ga 10^{-7} - 10^{-6} M_\odot \yr^{-1}$ (depending on the WD mass, e.g., \citealt{Hachisuetal1999, Boursetal2013}) nuclear burning takes place on the surface of the WD and prevents further accretion. We return to this point in section 
\ref{sec:Parasite}.

In the present study we explore a triple-star CEE where a giant star forces a tight WD-MS binary system to enter a tight-CEE while the three stars experience a triple-star CEE. Moreover, the interaction with the giant star, envelope and core, helps in unbinding the CEM of the WD-MS system, leaving behind a WD orbiting the core of the giant star, or else the WD and the core merge.  
As said, the WD-MS tight CEE is very different from a tight neutron star-MS CEE inside a giant star because the neutron star can launch jets (as \citealt{Soker2021TwoCEJSN} explored recently). It also differs from other triple-star CEE where two stars of the tight binary system merge with each other, launch jets, or the evolution disrupt the tight binary system (e.g.,  \citealt{SabachSoker2015, Hilleletal2017, Schreieretal2019, ComerfordIzzard2020, GlanzPerets2021}). 

The large fraction of massive stars in triple-star systems (and multiple-star systems in general; e.g., \citealt{MoeDiStefano2017}) motivates this study. Although there are rich varieties of processes that can take place before the binary system enters the envelope of a giant tertiary star (e.g.,  \citealt{IbenTutukov1999, deVriesetal2014, PortegiesZwartLeigh2019, Leighetal2020}), e.g., an accretion-induced merger, we here concentrate on the double-CEE of the WD-MS-AGB triple star system. 
 
As this is a novel scenario, we start by describing it in section \ref{sec:Scenario}. We calculate the binding energy of the tight-binary CEM and the tight WD-MS binary orbital energy in section \ref{sec:BindingEnergy}, and then (section \ref{sec:Parasite}) discuss the way the giant removes the CEM of the tight WD-MS binary system (a parasite CEE). We crudely estimate the rate of such events in section \ref{sec:rate}. We summarise in section \ref{sec:Summary}. 

% ==========================================================
\section{The parasite CEE scenario} 
\label{sec:Scenario}
% ==========================================================
 
\begin{figure*}%[ht]
 %\centering
%  \vskip -3.00 cm
 %\hskip -1.00 cm
\includegraphics[trim=0.0cm 0.0cm 0.0cm 1.0cm ,clip, scale=0.80]{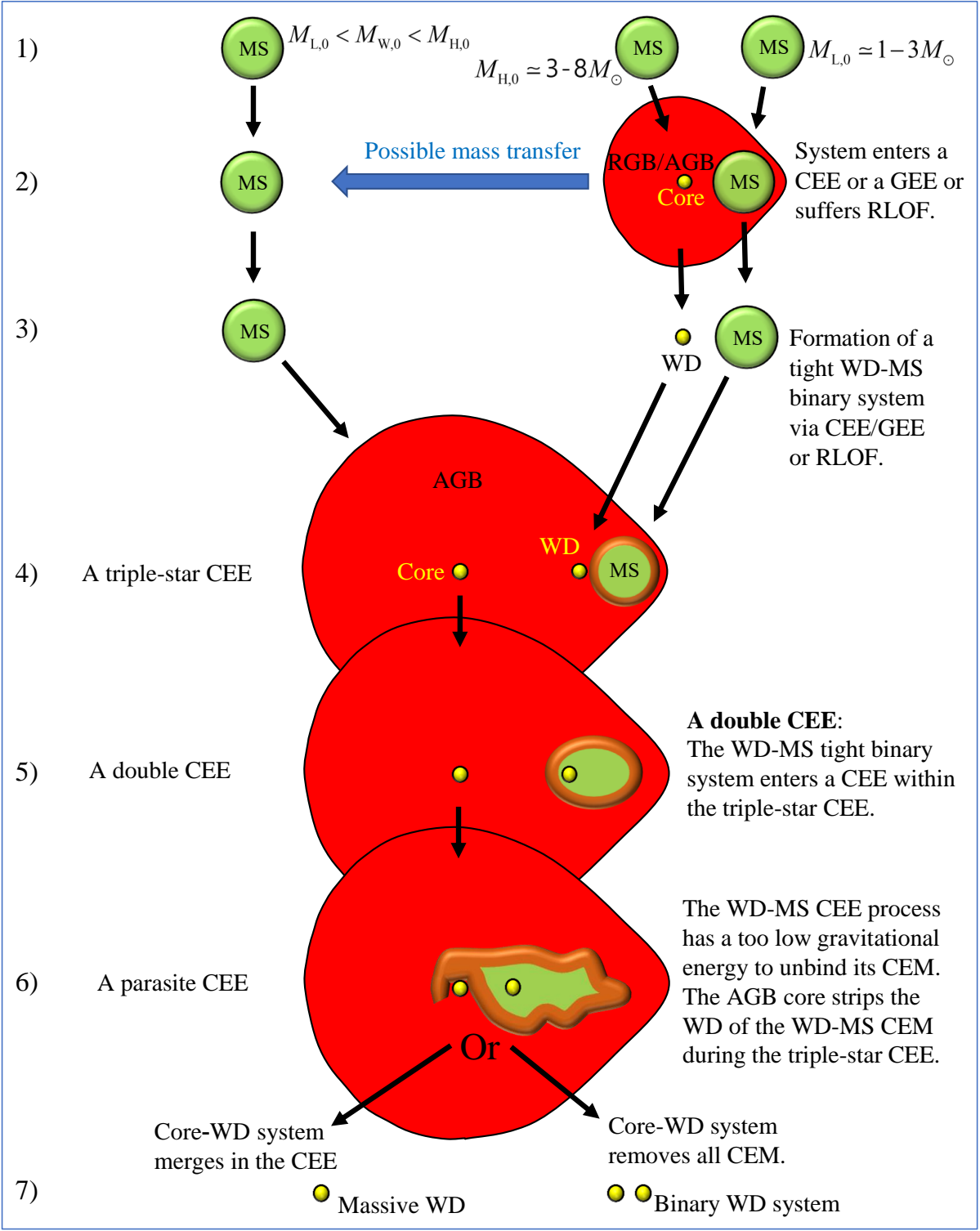}
% \includegraphics[]{CCSNIaFigure.pdf}\\
% \vskip -6.00 cm
\caption{A schematic diagram of the proposed double CEE scenario with a parasite CEE. 
Abbreviation: AGB: asymptotic giant branch; CEE: common envelope evolution; CEM: common envelope matter; GEE: grazing envelope evolution; MS: main sequence; RGB: red giant branch; RLOF: Roche lobe overflow; WD: white dwarf.   }
 \label{fig:SchematicScenario}
 \end{figure*}

In this section we discuss the different evolutionary phases of the double CEE that lead to the `parasite CEE'. We schematically present the scenario in Fig. \ref{fig:SchematicScenario}. We enumerate the phases by the row number in Fig. \ref{fig:SchematicScenario}.
We start by estimating the initial conditions on such an evolution, postponing a more accurate calculation of the evolution towards a triple CEE to a forthcoming paper. 
%FFFFFFFFFFFFFFFFFFFFFFFFFFFFFFFFFFFFFFFFF
  %[ht]
 % %FFFFFFFFFFFFFFFFFFFFFFFFFFFFFFFFFFFFFF
   
\textit{1. The initial triple-star system.} We require the most massive star in the triple to be in the inner (tight) binary system, and we require it to develop into a WD. As well, we require that at later phases the other star in the tight binary be massive enough to allow the WD to spiral inside it. So either it starts as massive enough, or it accretes mass from the most massive star. We also require the wide star to evolve next. The wide star can also accrete mass, in particular if the first CEE, that of the tight binary system, loses mass in the equatorial plane and the orbital plane of the triple is aligned with that of the inner binary.  

Overall, we estimate the requirement on the initial (zero age main sequence) mass of the highest-mass star to be $M_{\rm H,0} \simeq 3-8 M_\odot$. We also require the wide star, of initial mass $M_{\rm W,0}$, to be more massive than the other star in the tight binary system, of initial mass $M_{\rm L,0}$, such that it evolves to the AGB (or in some cases to the RGB) while the tight binary system is still a WD-MS one. Overall, we estimate the initial masses to be (see also Fig. \ref{fig:SchematicScenario})
\begin{eqnarray}
\begin{aligned}
3 M_\odot \la & M_{\rm H,0} \la 8 M_\odot  
\\ 
1 M_\odot \la  & M_{\rm L,0} \la 3 M_\odot  
\\ 
  M_{\rm L,0} < & M_{\rm W,0}  <  M_{\rm H,0}  ,
\label{eq:Minitial}
\end{aligned}
\end{eqnarray}
where the first two lines are for the tight (inner) binary system.

 The mass constraints bound the possible age of the system. The most massive star will form a WD on a time of $\simeq 10^7 - 5 \times 10^8 \yr$ from star formation, where the short time is for the upper mass limit. The duration of the next phase depends on the mass of the wide star after it acquires mass in the previous phase. This mass transfer can bring it to a mass of $M_{\rm W} \ga 3 M_\odot$, implying next phase evolution time of $\la 5 \times 10^8 \yr$. Overall, the typical time for the parasite CEE to take place is about $\simeq  10^8 - 10^9 \yr$ after the system was formed. As more massive stars are rarer, most parasite CEE cases will take place at $\approx 10^9 \yr$ after star formation.

 We next estimate the initial orbital separations. 
 
\textit{2. The first CEE of the inner binary.} 
The most massive star evolves to become a RGB star and, if no CEE takes place, later to an AGB star.  Envelope expansion and orbital shrinkage because of tidal interaction that steeply strengthens with the giant expansion, bring the binary system to a CEE (the first one in the triple system).  The remnant of this CEE is a tight WD-MS binary. 
For this to occur we require that the tight binary system both allows the most massive star to form a WD, but that the WD-MS tight binary ends as a detached binary with an orbital separation of $a_{\rm in} \sim 2-10 R_\odot$. Another constraint on the initial orbital separation comes from the requirement that the initial triple be stable, and the wide companion be able to engulf the tight WD-MS binary system. 
For the latter to occur, we estimate that the initial semi-major axis of the triple (the wide companion and the center of mass of inner binary) be  
$a_{\rm out,0} \simeq 0.5 - 3 \AU$. For the initial triple to be stable we estimate that (e.g., \citealt{AarsethMardling2001}) $a_{\rm out,0} \ga 3 a_{\rm in,0}$. Overall we estimate the initial semi-major axes for small eccentricities to be (depending on the mass ratio $q_0 \equiv M_{\rm W,0}/(M_{\rm H,0} + M_{\rm L,0}) \simeq 0.3- 0.7$) 
\begin{eqnarray}
\begin{aligned}
0.1 \AU \la      & a_{\rm in,0} \la 0.5 \AU   
\\ 
3 a_{\rm in,0}  \la & a_{\rm out,0} \la 3 \AU .  
\label{eq:Rinitial}
\end{aligned}
\end{eqnarray}

\textit{3. The formation of the tight (inner) WD-MS binary.} 
The descendant of the first CEE of the inner binary is a WD-MS tight binary system, i.e., the two stars orbit each other with a semi-major axis of $a_{\rm in} \approx 2-10 R_\odot$. In cases that the inner binary starts very close to each other, the system might evolve through a Roche lobe overflow (RLOF) and/or the grazing envelope evolution (GEE), in addition to or instead of the CEE.  We also expect in this case that the most massive star will lose its envelope on the RGB and the WD remnant will be a helium WD of $M_{\rm WD} \la 0.45 M_\odot$.  Future studies will determine the exact processes and the range of the required value of $a_{\rm in}$ and their dependence on the different other parameters of the triple. 

\textit{4. The early phase of the triple-star CEE.} 
During this phase the wide star expands and engulfs the tight binary system, forming triple-CEE, which is the second CEE of this system. Since we expect the wide star to be massive, $M_{\rm W} \ga 3 M_\odot$ at this phase, in most cases, but definitely not all, this phase will take place during the AGB phase of the wide star rather than during its RGB phase. 
The tight binary enters the AGB envelope and spirals in.
 In many other cases the wide star might engulf the tight binary system on its RGB. For that to take place the orbit of the wide star and the tight binary system should be relatively small, $\la 50-100 R_\odot$. This implies that the initial separation of the tight binary system was also small and the most massive star in the system lost its envelope also on the RGB, and so the WD in the tight binary system is a helium WD of $M_{\rm WD} \la 0.45 M_\odot$.   

The dynamical interaction with the core and the dynamical friction of the envelope intensify with the spiralling-in process. There are several possible outcomes of triple-CEE (e.g., \citealt{SabachSoker2015, ComerfordIzzard2020, GlanzPerets2021}). Here we take the case where the dynamical friction due to the interaction of the tight binary with the giant envelope shrinks the orbit of the tight binary (e.g., \citealt{GlanzPerets2021}). In addition, the accretion of mass by the MS star  causes it to expand. These two processes bring the MS star to engulf the WD and the binary system forms a tight CEE. This is the third CEE of the triple system. 
  
\textit{5. The tight-CEE inside the triple-star CEE.} 
Now the WD spirals-in inside the MS envelope. However, this process does not release enough orbital energy to unbind the WD-MS CEM, as we show in section \ref{sec:BindingEnergy}.

\textit{6. The parasite tight CEE: envelope removal by the triple interaction.} The release of orbital energy of the WD-MS system and the nuclear burning on the surface of the WD forms a giant-like structure that expands the envelope of the MS star. The ram pressure of the AGB envelope constantly removes the outskirts of this envelope, but only a small fraction of the CEM. Later, the WD-MS system comes close enough to the core for the core to tidally unbind the rest of the WD-MS CEM (section \ref{sec:Parasite}). 

\textit{7. Final outcome.}  The final system might be two closely orbiting WDs or the merger of the two WDs to form a massive WD. Nothing in the final stellar system will tell us that the system was a triple that had experienced the parasite CEE. 
However, within $\approx 10^5 \yr$ of total envelope ejection the system forms a planetary nebula. The triple is likely to leave a messy planetary nebulae, i.e., one that strongly depart from axial-symmetry, in particular if the orbital plane of the inner binary is not aligned with the triple orbital plane (e.g., \citealt{BearSoker2017, Jonesetal2019, RechyGarciaetal2020, GlanzPerets2021, Henneyetal2021}). Else, the system might explode as a (peculiar) SN Ia. We elaborate on these outcomes in section \ref{sec:Parasite}.
 
 Because the progenitors of the two WDs, the most massive star and the wide star, are of masses $> 2 M_\odot$ before they leave the MS, and because they experience CEE, the WD masses in most cases are in the range of $\simeq 0.6-0.65 M_\odot$. WD progenitor masses of $> 2 M_\odot$ imply that their large expansion to engulf the companion takes place mainly on the AGB when their core mass is already $\ga 0.6 M_\odot$. In other cases the most massive star engulfs its MS close companion while on the RGB. Another option, but less likely, is that the wide star might in some cases engulf the tight WD-MS binary while on the RGB instead than on its AGB. In these cases the WD masses are lower with $\la 0.45 M_\odot$ and mostly helium WDs. The CEE implies that the MS companion removes the AGB envelope (in case of an AGB star) before the core of the most massive star grows to a mass beyond $\simeq 0.6-0.65M_\odot$. The same holds for the core of the wide star when it becomes an AGB star and engulfs the tight binary system (in cases it does grow to an AGB). Overall, in most cases we expect the two WDs of the final remnant (if merger does not take place) to be CO WDs and to have masses in the range of $\simeq 0.6-0.65M_\odot$. In a minority of the cases one or two WDs might have a mass of $\la 0.45 M_\odot$, and likely be helium WD(s).   
   
 We can summarise the evolution as follows. The initial triple system is of two stars that orbit the most massive star in the system of mass $3 M_\odot \la M_{H,0} \la 8 M_\odot$. The lowest mass star in the system orbits in a relatively close orbit, i.e., the most massive star and the lowest mass star form a tight binary system. A third star, the wide star, orbits in a wider orbit (masses and orbits in equations \ref{eq:Minitial} and \ref{eq:Rinitial} respectively). The product of the last CEE (line 6 in Fig. \ref{fig:SchematicScenario}) is a binary system composed of the core of the wider star and the WD remnant of the most massive star. The WD-core system either merges to form one massive WD or ends as a close WD-WD binary system.

% ==========================================================
\section{The energetic of the tight WD-MS CEE} 
\label{sec:BindingEnergy}
% ===========================================================

\begin{figure}%[ht]
 %\centering
 %\vskip -3.00 cm
 \hskip -2.00 cm
\includegraphics[trim=1.1cm 6.5cm 0.0cm 8.3cm ,clip, scale=0.6]{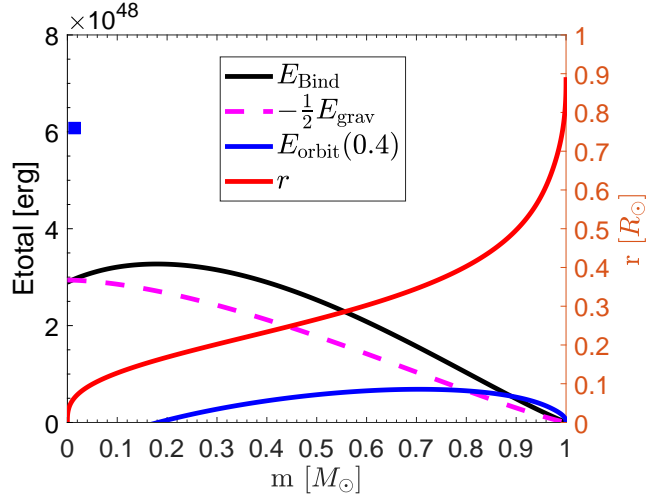}
 \vskip -1.00 cm
\caption{The relevant quantities as function of mass coordinate for a MS model of $1M_\odot$ at an age of $10^8 \yr$. 
We plot the binding energy of the envelope residing above mass $m$, $E_{\rm bind}$ (black line), the radius $r$, the absolute value of  half the gravitational energy of the envelope residing above $m$, $-(1/2)E_{\rm grav}$, and the orbital energy that a WD companion of mass $M_{\rm WD}= 0.4 M_\odot$ releases as it spirals-in from the MS surface to mass coordinate $m$ and removes the envelope.  The square on the left axis marks the energy that the WD would have released had it spiralled-in to the center without modifying the MS structure (equation \ref{eq:Ecenter}). }
 \label{fig:1Mo0.4WDmodel}
 \end{figure}

\begin{figure}%[ht]
 %\centering
% \vskip -1.50 cm
%\hskip -1.50 cm
% Trim: Left; down; right; up
\includegraphics[trim=3.5cm 8.5cm 0.0cm 8.3cm ,clip, scale=0.6]{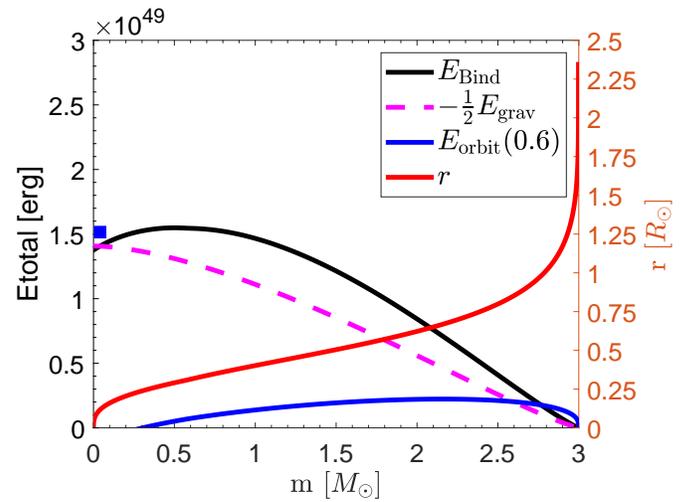}
% \includegraphics[]{CCSNIaFigure.pdf}\\
%\vskip -1.50 cm
\caption{Similar to Fig. \ref{fig:1Mo0.4WDmodel} but for a model of $3 M_\odot$ and a WD of $M_{\rm WD}= 0.6 M_\odot$.   }
 \label{fig:3Mo0.6WDmodel}
 \end{figure}

We simulate the evolution of two stellar models with zero age main sequence masses of $1M_\odot$ and $3M_\odot$ to an age of $t_{\rm model} = 10^8 \yr$, as the other two stars in the triple are more massive and evolve on short time scales. We use   
\textsc{mesa} (Modules for Experiments in Stellar Astrophysics; version 10398; \citealt{Paxtonetal2011, Paxtonetal2013, Paxtonetal2015, Paxtonetal2018, Paxtonetal2019}) following the example of $1M~pre~ms~to~wd$ of \textsc{mesa}.  We present the relevant quantities for the two models in Fig. \ref{fig:1Mo0.4WDmodel} and Fig. \ref{fig:3Mo0.6WDmodel}, respectively.
%FFFFFFFFFFFFFFFFFFFFFFFFFFFFFFFFFFFFFFFFF
  %[ht]
 % %FFFFFFFFFFFFFFFFFFFFFFFFFFFFFFFFFFFFFF
%FFFFFFFFFFFFFFFFFFFFFFFFFFFFFFFFFFFFFFFFF
  %[ht]
 % %FFFFFFFFFFFFFFFFFFFFFFFFFFFFFFFFFFFFFF

We check the portion of the MS envelope that the WD might remove as it spirals in inside the MS envelope. For that we compare the binding energy of the envelope mass that resides above mass coordinate $m$ at radius $r(m)$ and up to the MS surface $M_{\rm MS}$ at radius $R_{\rm MS}$. The binding energy is the sum of the gravitational energy $E_{\rm grav}(m)$ of that zone and its internal energy $U(m)$
\begin{equation}
E_{\rm bind} (m) = - \int^{M_{\rm MS}}_m 
\left[ -\frac {G M_r}{r(m)} + e(m) \right] dm , 
\label{eq:Ebind}
\end{equation}
where $M_r$ is the MS mass inner to radius $r$, and $e(m)$ is the internal energy per unit mass. 
We plot $E_{\rm bind}$ by a solid black line.  
For comparison purposes we also present (dashed-pink line) the quantity $-(1/2)E_{\rm grav}(m)$.

We  calculate the gravitational energy that the WD-MS system releases due to the spiralling-in process from initial radius $a_0=R_{\rm MS}$ to radius $a=r(m)$. We calculate this energy under the assumption that the WD manages to eject all MS envelope residing outside radius $r(m)$, i.e., all mass outside mass coordinate $m$. This energy due to the shrinking orbit with mass removal is 
\begin{equation}
    E_{\rm orbit} = \frac{G M_a M_{\rm WD}}{2 a} - \frac{GM_{\rm MS}M_{\rm WD}}{2R_{\rm MS}}, 
\label{eq:Eorbit}
\end{equation}
where $M_a$ is the MS mass inner to $r=a$. 
We plot $E_{\rm orbit}$ by a solid-blue line for each of the two models. For the $1 M_\odot$ MS model we plot $E_{\rm orbit}(0.4)$ for $M_{\rm WD}=0.4 M_\odot$, while for the $3 M_\odot$ MS model we plot $E_{\rm orbit}(0.6)$ for $M_{\rm WD}=0.6 M_\odot$.

From Figs. \ref{fig:1Mo0.4WDmodel} and \ref{fig:3Mo0.6WDmodel} we learn that the WD can remove only a fraction of $f_{\rm WD,ej} \la 10- 20\%$ of the MS mass. For a more reasonable CEE efficiency of $\alpha_{\rm CEE} <1$ we expect even less and take $f_{\rm WD,ej} \la 10\%$. 

The energy $E_{\rm orbit}$ is the difference between the WD-MS orbital energy when the WD orbits at the surface of the MS star to the orbital energy when the semi major axis is $a<R_{\rm MS}$ under the assumption that the entire envelope that initially resided at $r>a$ has been already ejected. The other extreme is that the envelope structure does not change at all. In this case the WD releases energy as it spirals-in all the way to the center. The total energy that the WD releases as it spirals-in from a Keplerian orbit at the surface of the MS star, with an initial kinetic energy of $E_{\rm k,MS}=GM_{\rm MS} M_{\rm WD}/2R_{\rm MS}$, to be at rest at the center of the undisturbed MS star is 
\begin{eqnarray}
\begin{aligned}
E_{\rm a=0} & = \int ^0 _{R_{\rm MS}} 
\left[ - \frac {G M_{\rm MS}(r) M_{\rm WD}}{r^2} dr  \right] + E_{\rm k,MS} 
\\ & =   \int_0 ^{M_{\rm MS}} 
\frac {G M_{\rm WD}}{r} d m   - \frac{1}{2} \frac{ G M_{\rm MS} M_{\rm WD}}{R_{\rm  MS}} .
\label{eq:Ecenter}
\end{aligned}
\end{eqnarray}

If we to calculate envelope ejection with this energy we must consider that now the envelope has to overcome the gravitational potential well of the WD, and we are back to the conclusion that the binary WD-MS system can at best expel a small fraction of the envelope mass by itself, or more likely it will inflate somewhat the WD-MS CEM. For comparison we do mark $E_{\rm a=0}$ on the graph with a blue square symbol. 

When the WD starts to accrete mass a nuclear burning starts on its surface. For the WD masses we take here the nuclear power would be $L_{\rm WD,nuc} < 10^4 L_\odot$ (e.g., \citealt{Hachisuetal1999, Boursetal2013}). If the WD-MS system reaches the core of the AGB star with a time of $t_{\rm CEE} <100 \yr$, the nuclear energy would add an energy of $E_{\rm WD,nuc} < 10^{47} \erg$. This energy does not change our conclusion.  
In an isolated WD-MS binary that merges, over hundreds to thousands of years the nuclear energy would inflate the envelope to form a giant star. However, in the present triple-star double-CEE other processes take place, as we describe next. 

% ==========================================================
\section{Parasite envelope removal} 
\label{sec:Parasite}
% ===========================================================
 
The triple now is in a phase of double-CEE. 
In the inner CEE a WD spirals-in inside the envelope of a MS star, and this tight binary system spirals-in inside the envelope of a giant star. As we showed in section \ref{sec:BindingEnergy} (Figs. 2 and 3) the WD can at most eject a fraction of $f_{\rm WD,ej} \la 10\%$  of the MS mass. 
The rest will stay bound, possibly in a somewhat inflated envelope. 
   
If it was not for the giant star, an isolated WD-MS system would 
have evolved to a giant star (RGB for a helium WD and an AGB for a CO WD) by the nuclear burning on the surface of the WD. It would have taken this giant about $10^5 - 10^7 \yr$ to lose its envelope in a wind. 
 However the triple CEE with the giant and its core removes the envelope on much shorter timescales. 
There are two branches for the future evolution of the triple.

% ==========================================
\subsection{WD-core merger} 
\label{subsec:Merger}
% ==========================================
 
This evolutionary branch applies for cases of massive and compact giants, like massive RGB or early AGB giants, that have a large binding energy of \begin{equation}
E_{\rm env,G} \ga \alpha_{\rm CEE} \frac {G M_{\rm core} M_{\rm WDMS}} {1 R_\odot} ,
\label{eq:Eenvmer}
\end{equation}
where $M_{\rm core}$ is the core mass, $M_{\rm WDMS}$ is the mass of the WD-MS inner binary system, and $\alpha_{\rm CEE}$ is the CEE efficiency parameter.
At an orbital separation of $a_m \simeq 1 R_\odot$ the WD-MS suffers RLOF, and the system turns to a WD-core binary system inside the dense envelope of the destroyed MS star. We expect the WD-core system to spiral-in towards each other on a time scale of hours to days, which is about ten times the dynamical time of the system at this stage. Gravitational drag with the envelope and tidal interaction between the core and the WD will most likely bring the WD-core to merge, ending the evolution with the formation of a massive and rapidly rotating WD. If the mass of the merger remnant is close to the Chandrasekhar mass limit, this WD might explode in the near or far future as a (peculiar) SNe Ia in the frame of the core degenerate scenario of SNe Ia. 

We emphasise that it is the WD-core interaction that removes the CEM of the WD-MS system. Namely, the `parasite' WD-MS CEE `uses' the energy of the triple CEE to unbind its envelope.   

% ==========================================
\subsection{AGB envelope ejection and a WD-core evolution} 
\label{subsec:AGBejection}
% ==========================================

In these cases the spiralling-in process of the triple system, i.e., the WD-MS tight CEE binary and the core of the AGB star, releases sufficient energy to eject the AGB envelope before the WD-MS system reaches the AGB core and suffers a rapid RLOF. Over a long time the nuclear burning on the surface of the WD inflates the WD-MS envelope to suffer RLOF to the core, reestablishing a diluted AGB envelope, i.e., the triple star CEM. The further spiralling-in of the WD and the core towards each other removes the diluted reestablished AGB envelope. 

The inflation timescale of the WD-MS envelope determines the evolution time. The timescale is therefore the thermal timescale of the WD-MS system, i.e., the binding energy of the WD-MS envelope divided by the nuclear luminosity of the WD
\begin{eqnarray}
\begin{aligned}
\tau_{\rm 2WDs} & \approx \frac {E_{\rm bind}}{L_{\rm WD,nuc}} = 5000  
\left( \frac {E_{\rm bind}}{3\times10^{48} \erg} \right)
\\ & \times
\left( \frac {L_{\rm WD,nuc}}{5 \times 10^3L_\odot} \right)^{-1}
\yr .
\label{eq:Tau2WDs}
\end{aligned}
\end{eqnarray}

Practically, we have here a CEE of a WD-core system in an evolving envelope. The evolution might leave two WDs, or, less likely, they might merge as a result of the angular momentum that the wind carries from the system. 
   
Again, it is the interaction in the triple system, the WD-MS with the core of the giant, inside the reestablished AGB envelope that removes the envelope of the WD-MS system.

% ==========================================================
\section{Event rate} 
\label{sec:rate}
% ===========================================================

\cite{Leighetal2020} conduct a population synthesis study and conclude that $f_3 \approx 1-5 \%$ of triple star systems produce a tight WD–MS inner binary accreting from an evolved wider companion. From the orbital periods distribution that they present in  their figure 7  we crudely estimate that about half of these systems are sufficiently close,  i.e., $a\la 4 \AU$,  to possibly  enter a CEE, $f_{\rm CEE} \approx 0.5$. 
In $f_q\simeq 80 \%$ of these cases the MS star in the tight binary is more massive than the WD. 

\cite{Leighetal2020} discuss a merger due to RLOF mass transfer from the wide star to the tight binary system. They did not study with their population synthesis the possibility of a merger during a CEE. It is quite uncertain what would be the fraction of the MS-WD systems that merge during the triple CEE. We expect this fraction not to be too small, $f_{\rm mer} \approx 0.1-0.5$, because we expect the MS to expand upon accreting mass from the AGB envelope.

We also cannot allow a too low-mass wide star if we require it to unbind the CEM of the WD-MS system. Crudely, $M_{\rm W,0} \ga 2 M_\odot$, which, by the initial mass function, reduces the relevant number of systems by about a factor of $f_w \simeq 0.3$. 
  
Overall, we very crudely estimate the fraction of parasite CEE out of all evolved triples to be 
\begin{equation}
f_{\rm Parasite} \approx f_3 f_{\rm CEE} f_q f_{\rm mer} f_w \approx 10^{-3}.
\label{eq:fraction}
\end{equation}
 
\cite{BearSoker2017} estimate that $\approx 13–21$ of PNe have been shaped by triple stellar systems. They could classify about 150 planetary nebulae that might have been shaped by triple star interaction. The chance that one of these have been shaped by the parasite CEE is $\approx10\%$.  

More promising is the possibility that one of the peculiar supernovae in the coming years will be via this channel. If the merger of the WD with the core leads to a thermonuclear explosion in the frame of the core degenerate scenario, then this supernova might be classified as a peculiar one due to the massive hydrogen-rich envelope and/or massive hydrogen circumstellar matter,  together with a large amount of nickel 56 that is synthesised in the explosion. 

Although the parasite CEE is a very rare triple evolutionary channel, so are many other triple evolutionary channels. We should explore all these rare channels.

% ==========================================================
\section{Summary } 
\label{sec:Summary}
% ===========================================================
 
We proposed and explored the basic properties of a rare triple star evolutionary channel that we term \textit{parasite CEE}. In this evolutionary channel an evolved triple star system enters a double-CEE evolution (fifth row of Fig. \ref{fig:SchematicScenario}), where a tight WD-MS system experiences a CEE while this binary and a wide giant (most likely an AGB star) experience a triple CEE. 

The wide AGB star plays two roles. Firstly, it induces the formation of the WD-MS tight CEE. An isolated WD-MS binary system is unlikely to enter a CEE during the MS phase. However, after this tight (inner) binary system enters the AGB envelope, the MS star accretes mass and swells to engulf the WD. In addition, the gravitational drag of the WD and MS components of the tight binary as they orbit each other inside the AGB envelope causes their mutual orbit to shrink. These two effects might bring the WD-MS binary to enter a CEE with each other. 
 
The second role of the giant, and in particular its core, is to remove the CEM of the WD-MS binary. A WD spiralling-in inside the MS envelope can at most unbind a fraction of $f_{\rm WD,ej} \la 10\% - 20\%$ of the MS mass (Figs. \ref{fig:1Mo0.4WDmodel} and \ref{fig:3Mo0.6WDmodel}). 
It is the interaction of the core with the WD-MS envelope, either by WD-core merger or by RLOF, that removes the WD-MS CEM. Namely, the triple CEE supplies the energy to remove the binary CEM. Hence our terminology of a \textit{parasite CEE.}  
  
The CEE evolution ends in one of two ways, either a WD-core merger or the formation of a double WD binary system. The merger process forms a massive WD, that if close to the Chandrasekhar mass limit might explode as a (peculiar) SN Ia in the frame of the core degenerate scenario of SNe Ia. The second outcome can also lead to a SN Ia if the two WDs merge on a later time, either in the frame of the double degenerate scenario or in that of the double detonation scenario. 

We very crudely estimated the rate of the parasite CEE to be a fraction of $\approx 0.001$ out of all evolved triple stars (equation \ref{eq:fraction}). 

 An extremely rare evolutionary route that we did not consider up to now is when instead of the MS star in the tight binary system there is a star that has just left the MS, i.e., a Hertzsprung gap star. This star is developing a helium core and a more extended hydrogen-rich envelope. We expect that in these minority of cases of parasite CEE the WD will be able to eject most of the hydrogen-rich envelope. However, it will most likely merge with the core of the Hertzsprung gap star because the Hertzsprung gap star is still compact. The merger of the helium core with the CO WD forms a star like an R Coronae Borealis star (e.g., \citealt{Munsonetal2021}). The merger product is a CO core and a helium-rich envelope. Only an interaction with the core of the giant star will remove the helium-rich envelope of the merger product. In a sense, although the CEE evolution of the WD with the Hertzsprung gap star might eject most or even all its hydrogen-rich envelope, we expect that a parasite CEE will still take place with respect to the helium-rich envelope.   

On a broader scope, we can group the very rare parasite CEE with many other very rare triple star evolutionary channels. The very rich variety of evolutionary channels that evolve triple star systems can take implies that many of these channels are very rare. Over the time we should explore these very rare evolutionary channels. We can expect that every year existing and future sky surveys will observe several transients that we will be able to attribute to one very rare triple evolutionary channel or another, like the possibility for a peculiar SN Ia of the parasite CEE.      

% ===================================================
\section*{Acknowledgments}
% ===================================================
 We thank an anonymous referee for helpful comments and suggestions. 
This research was supported by a grant from the Israel Science Foundation (769/20). 

%%%%%%%%%%%%%%%%%%%%%%%%%%%
\textbf{Data availability}
The data underlying this article will be shared on reasonable request to the corresponding author.  
%%%%%%%%%%%%%%%%%%%%%%%%%%%

% %%%%%%%%%%%%  References %%%%%%%%%%%%%%%%%%%%%

%%%%%%%%%%%%%%%%%%%%%%%%%%%%%%%%%%%%%%%%%%%%%%%%%%

\label{lastpage}
\end{document}